# Stationary and moving breathers in (2+1)-dimensional O(3) nonlinear σ-model

F. Sh. Shokirov

*S. U. Umarov Physical-Technical Institute of Academy of Sciences of the Republic of Tajikistan, Aini Avenue 299/1, Dushanbe*

**Abstract.** The formation and evolution of stationary and moving breather solutions in (2+1)-dimensional O(3) nonlinear σ-model are investigated. The analytical form of oscillating solutions for (2+1)-dimensional sine-Gordon equation, which evolve to periodic (breather) radially symmetric solutions is determined. On the basis of the found solutions by adding the rotations to the A3-field vector in isotopic space of $S^2$, the solutions for the O(3) nonlinear σ-model are obtained. By numerical study of the solutions dynamics their stability in a stationary and a moving state for quite a long time (45000 cycles), although in the presence of weak radiation is shown.

**I. Introduction**

The study of self-localized in space and periodic in time breather type solutions (bions, breathers, doublets, pulsons etc.) of nonlinear field theory models, consisting of two antiphase-solitons, has attracted many researchers [1-8]. Investigation of the processes of formation of spatially homogeneous oscillation modes, which can be interpreted as the classic model of composite particles (e.g., mesons field) is an urgent task in the soliton theory. Determination of the formation conditions of stable particle-like solutions is one of the key questions at the study of nonlinear field theory models, which admit localized (soliton) solutions. A special class of practical problems in the soliton theory is the study of oscillating solutions, coherent structures that have their own dynamics of internal degrees of freedom.

In the present work considered a method for solving this problem proposed by the authors of [1], who conducted a study of processes of formation and evolution of the breather solutions for two-dimensional sine-Gordon equation (SG)

$$u_{tt} - u_{xx} - u_{yy} + \sin u = 0, \tag{1}$$

in the Lagrangian and Hamiltonian approaches. In particular, based on the trial function

$$u(x,y,t) = -4\,arctg[a(t)\sin(v(t))\,\text{sech}(\lambda(t)x)\,\text{sech}(\lambda(t)y)], \tag{2}$$

$$a(t) = \frac{\lambda(t)}{\sqrt{1-\lambda(t)^2}},$$

by authors [1] in the Hamiltonian approach the numerical and asymptotic solutions of the equation (1) were obtained by in the form of a stable stationary and moving breathers solitons.

In the present paper continue the study, carried out in [1] in the Lagrangian approach. In particular, by averaged the Lagrangian

$$L = \frac{1}{2}(u_t^2 - u_x^2 - u_y^2) - 1 + \cos u, \tag{3}$$

for the equation (1) with respect to the fast phase $v(t)$ and by solution the integral equation [1]

$$\mathcal{L} = \frac{1}{2\pi}\int_0^{2\pi} L\,dv \tag{4}$$

the expressions for the phase variables $\lambda(t)$ and $\nu(t)$ test solutions (2) are derived, which are given in a second section of paper. On the basis of the found solutions of SG equation (1), and by adding specially selected perturbation of A3-field vector in isotopic space of $S^2$ [9-12] the solutions of (2+1)-dimensional O(3) nonlinear σ-model (NSM) are obtained. In the third and fourth sections of the paper are given the results of numerical simulations [13,14] of the found breathers of equation (1) (as test tasks) and (2+1)-dimensional O(3) NSM in stationary and moving cases respectively. The obtained in the framework of the O(3) NSM the formed stable periodic breather solutions, in the evolution are exempt from the excess energy in the form of radiated radially symmetric linear perturbation waves. These waves is absorbed at the boundaries of simulation area by the specially boundary conditions [9-12,15,16]. In the last section of the paper discusses the properties of the obtained results and their possible application to the study of other tasks.

## II. Theoretical calculations

In the present paper, the averaged density of the Lagrangian (3) respect to the fast phase $\nu(t)$ is obtained in the form of

$$\mathcal{L} = \Xi(\lambda)\lambda_t^2 + \Omega(\lambda)\varphi_t^2 - \Theta(\lambda), \qquad (5)$$

where

$$\Xi(\lambda) = \frac{2}{(1-\lambda^2)^{3/2}}, \quad \Omega(\lambda) = \frac{4\lambda^2}{\sqrt{1-\lambda^2}}, \quad \Theta(\lambda) = \frac{48}{\lambda}\psi(\lambda)\operatorname{arth}(\lambda) - 32,$$

$$\lambda = \lambda(t), \nu = \nu(t).$$

In [1] it was noted that the dynamics of the trial function (2) can be described by investigating the properties of two phase parameters: $\lambda(t)$ and $\nu(t)$, which in this case are defined as follows:

$$\lambda(t,D) = \pm \operatorname{th}\left(\sqrt{3}D\sqrt{\frac{-1\pm\sqrt{1+\sqrt{3}\frac{(D^2-288)}{3D\psi(\lambda)}}t}{\psi(\lambda)D^2 - 288}}\right) \qquad (6)$$

$$\nu(t,D) = -\frac{\sqrt{1-\lambda^2} + \lambda\,\psi(\lambda)\arcsin(\lambda)}{4\lambda}D, \qquad (7)$$

where $D = D(\nu_{t_0})$ – a constant (conserved quantity [1]), defined in this case by initial speed of (2).

Recall that the Euler-Lagrange equation of the investigated in the present work O(3) NSM for the anisotropic case is [10-12,15,16]

$$2\partial_\mu \partial^\mu \theta + \sin 2\theta \left(1 - \partial_\mu \varphi \partial^\mu \varphi\right) = 0, \qquad (8)$$

$$2\cos\theta\,\partial_\mu\varphi\partial^\mu\varphi + \sin\theta\,\partial_\mu\partial^\mu\varphi = 0,$$

where $\mu = 0,1,2$; $\theta(x,y,t)$ and $\varphi(x,y,t)$ – are Euler angles, associated with isospin parameters of model (8) as follows:

$$s_1 = \sin\theta\cos\varphi, \quad s_2 = \sin\theta\sin\varphi, \quad s_3 = \cos\theta,$$

$$s_i s_i = 1, \quad i = 1,2,3.$$

Equations (8) in a special parameterization ($2\theta$), in meridian section ($\varphi_0(x,y,t) = const$) of isotopic space $S^2$ are reduced [9-12] to the equation (1) in parameterization $2\theta$:

$$2\Box\theta = -\sin 2\theta. \tag{9}$$

Thus, as an initial approximation can be used expression for trial function (2) of model (1) and, respectively, introducing into it a specially selected perturbation in form

$$\varphi(x,y,t) = \varphi_0(x,y,t_0) + \omega\tau, \quad \omega \neq 0,$$

by solving the Cauchy problem, can get a new numerical solutions of the model (8) [9-12].

For field functions $s_i$ ($i = 1,2,3$) we have

$$s_1 = -\frac{2\xi}{1+\xi^2}\cos\varphi, \quad s_2 = -\frac{2\xi}{1+\xi^2}\sin\varphi, \quad s_3 = \frac{1-\xi^2}{1+\xi^2}, \tag{10}$$

where

$$\xi(x,y,t) = \frac{\lambda(t)}{\sqrt{1-\lambda(t)^2\psi}}\frac{\sin(\nu(t))}{\cosh(\lambda(t)x)\cosh(\lambda(t)y)},$$

$$\varphi = \omega\tau, \quad \omega \neq 0.0.$$

In the present paper, is applied an algorithm to the numerical scheme developed in [9,15,16], where we have used the properties of stereographic projection: the points of the upper hemisphere ($s_3 > 0$) are projected onto the tangent complex plane, passing through the "north pole"; points of the lower hemisphere ($s_3 < 0$) are projected onto a tangent plane, passing through the "south pole" of Bloch sphere $S^2$. At points of the "equator" ($s_3 = 0$) is made in a special way "crosslinking" the solution, and thereby carried out the bijective projection (compactification $S^2 - R^2_{comp}$):

$$z = x + iy = \frac{s_1 + is_2}{1 \pm s_3} = \text{tg}\frac{\theta}{2}e^{i\varphi}$$

of the all points of the complex plane z (including the $(x,y) = \infty$) and the sphere $S^2$:

$$s_i s_i = 1, (i = 1,2,3).$$

Was compiled and applied the three layers explicit difference schemes [9-16] with the approximation error of the second order of accuracy with respect to time and coordinates of $O(h^2 + \tau^2)$. Taking into account the effects of the boundary absorbing conditions and radiation of formed breather the excess energy in the form of linear perturbation wave, the energy integral of obtained models, preserved with good accuracy: $En_{loss} < 5.15\%$ in during of 45000 iterative cycles ($t \in [0.0, 270.0]$). A good agreement between the numerical and analytical calculations is obtained.

### III. Stationary breathers

In the first stage as test models the stable stationary numerical breathers (10) of models (9) (in frame of the model (8), with $\omega = 0.0$) is obtained, which was originally not being radially symmetrical, evolving to a radially-symmetrical form. These properties are investigated in detail in [1]. However, unlike the results of [1] in our paper the evolution of breather solutions to the radially-symmetrical form it occurs periodically (Fig.1, see also, Appendix, Fig.A.1 and Fig.A.2).

In Fig.1a and Fig.1b are shown the initial ($t = 0.0$) and the final state ($t = 270.0$) of energy density (DH) of breather solution (10) in model (8) at $\omega = 0.0$. The energy density

(DH) is calculated based on the Hamiltonian of O(3) NSM (in isospin parameterization) [10-12,15,16]:

$$\mathcal{H}_{Ip} = \frac{1}{2}[(\partial_0 s_a)^2 + (\partial_1 s_a)^2 + (\partial_2 s_a)^2 + (1 - s_3^2)],$$
$$s_a s_a = 1, \quad a = 1,2,3.$$

In Fig.1c shows the evolution of the energy density (DH) of breather soliton (10) in a planar section $(x, y_0)$, $y_0 = 0.0$. However, the illustration of contour projections in Fig.1b indicates that at the end of the simulation time ($t = 270.0$) the breather does not have a radially symmetrical shape. As noted above, transition to the radially-symmetrical form in this case is periodic. Note the change in the energy density values (DH) of the central part of the breather (point $x_0, y_0$), obtained by numerical simulations and analytical methods shown in Fig.1d. In this case, there is a very good agreement between analytical and numerical results. In Fig.1d shown that during the simulation time ($t \in [0.0, 270.0]$) amplitude values of the center point probation can be divided into repetitive wave packets (in this case observed a 7, with different duration $t \in (30.0, 45.0)$). In each wave packet the breather solution (10) is initially not being radially symmetrical (Fig.1a) is evolving to a radially-symmetrical form, but at the end of each packet loses this property (see, e.g., Fig.1b, in more detail in Fig.A.1 of Appendix).

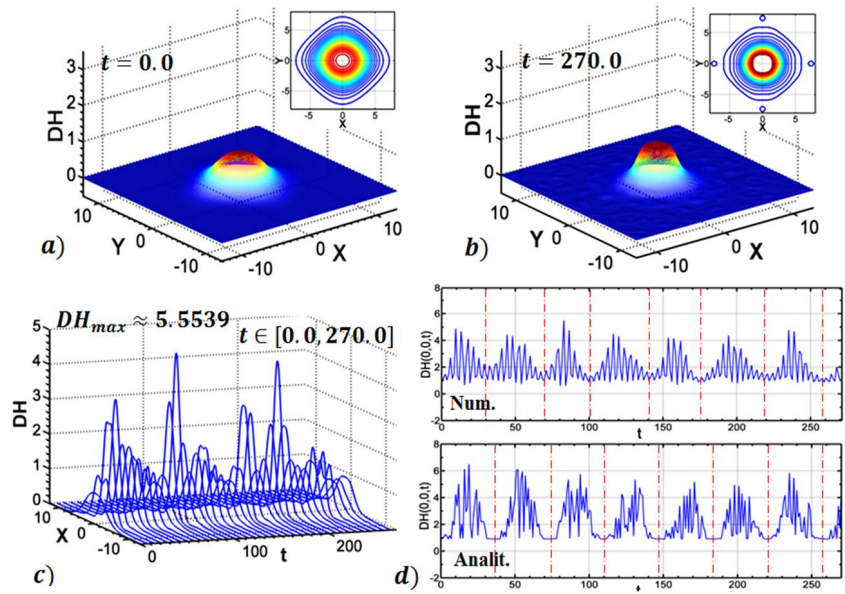

**Fig. 1.** Evolution of the breather solutions (10) of model (8) at $\omega = 0.0$ (SG). The energy density (DH) and its contour projection at $D(v_{t_0}) = \text{sh}(4.38)$: **a)** $t = 0.0$; **b)** $t = 270.0$; **c)** in the planar section $(x, y_0)$, $y_0 = 0.0$; **d)** $DH(0,0,t)$, obtained by numerical simulation (Num.) and analytically (Analit.). Simulation time: $t \in [0.0, 270.0]$.

Fig.1c shows the evolution of the solution profile of DH in planar section, but the detailed illustration of the first and beginning of the second wave packets (at $t \in [0.0, 42.0]$) is given in the Appendix (Fig.A.1). The contour projections of DH on the Fig.A.1 show that up to $t \approx 9.6$ (b) breather gets a radially symmetrical form. At the end of the first and beginning of the second wave packets the breather begins to lose a radially symmetric structure; this process is clearly visible at $t \approx 38.4$ and $t \approx 39.6$ (Fig.A.1f).

For the test models (described in Fig.1 and Fig.A.1) by adding the rotations to the A3-field vector (in this case: $\omega = 0.5$) in the isotopic space $S^2$ were obtained (Fig.2) the stability breather of the (2+1)-dimensional O(3) NSM. Fig.2 shows the initial (a) and final states (b) of energy density (DH) of numerical breathers (10) of O(3) NSM (8) at $\omega = 0.5$; simulation time $t \in [0.0, 270.0]$. On Fig.2c shows the DH evolution of the breather (10) of model (8) in a planar section $(x, y_0)$ $(y_0 = 0.0)$. Changes in the energy density values (DH) of the central part of the breather (point $x_0, y_0$), obtained by numerical simulation (Num.) and analytically (Analit.) are shown in Fig.2d. In this case, as in the previous example (Fig.1d at $\omega = 0.0$), there is a good concordance of analytical and numerical calculations. Fig.2d shows that during the whole simulation time ($t \in [0.0, 270.0]$) values of the center point of the amplitude (DH) can be divided into repetitive wave packets (in this case, conditionally 3 packets of waves of different duration $t \in (100.0, 130.0)$).

In each wave packet the breather solution is initially not being radially symmetrical (Fig.2a) is evolving to a radially-symmetrical form, but at the end of each packet loses this property (see, e.g., Fig.2b, in more detail in Fig.A.2 of Appendix).

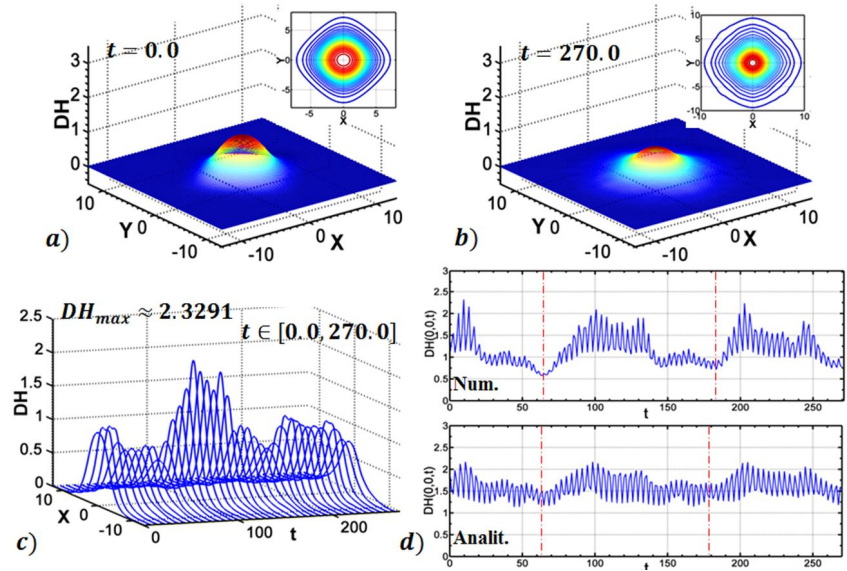

**Fig. 2.** Evolution of the breather solutions (10) of model (8) at $\omega = 0.5$. The energy density (DH) and its contour projection at $D(v_{t_0}) = \text{sh}(4.38)$: **a)** $t = 0.0$; **b)** $t = 270.0$; **c)** in planar section $(x, y_0)$, $y_0 = 0.0$; **d)** $DH(0,0,t)$, obtained by numerical simulation (Num.) and analytically (Analit.). Simulation time: $t \in [0.0, 270.0]$.

Detailed illustration of the evolution of the energy density (DH) of breathers (10) of O(3) NSM in the first $t \approx [0.0, 64.8]$ and in the beginning of the second $t \approx [73.2, 97.2]$ wave packets are shown respectively on Fig.A.2a-e and Fig.A.2f of Appendix. The contour projections of (DH) on Fig.A.2 show that up to the time of $t \approx 20.4$ (c) breather is gradually taking a radially-symmetrical form. At the end of the first and beginning of the second wave packets breather begins to lose a radially symmetrical structure. This process continues for a long time and is clearly visible on Fig.A.2.d-f at $t \approx [34.8, 87.6]$. Toward the middle of the second wave packet breather regains a radially symmetrical shape (Fig.A.2f at $t \approx 97.2$). Note that the energy density values (DH) of breathers (10) in O(3) NSM (8) in the presence of rotation of A3-field vector $\omega \neq 0.0$ ($DH_{max} \approx 2.3291$ in the case of $\omega = 0.5$) is much smaller relative to the case $\omega = 0.0$ ($DH_{max} \approx 5.5539$). Note also that breathers (10) of O(3) NSM (8) ($\omega \neq 0.0$) have more energy (En) relative breathers of SG equation (9) (within the O(3) NSM at $\omega = 0.0$).

Fig.3a-d shows the comparison of the energy density (DH) of numerical breathers (10) of O(3) NSM (8) and SG equation (9) in the initial stage of their evolution (in central planar section) at $t \in [0.0, 13.2]$.

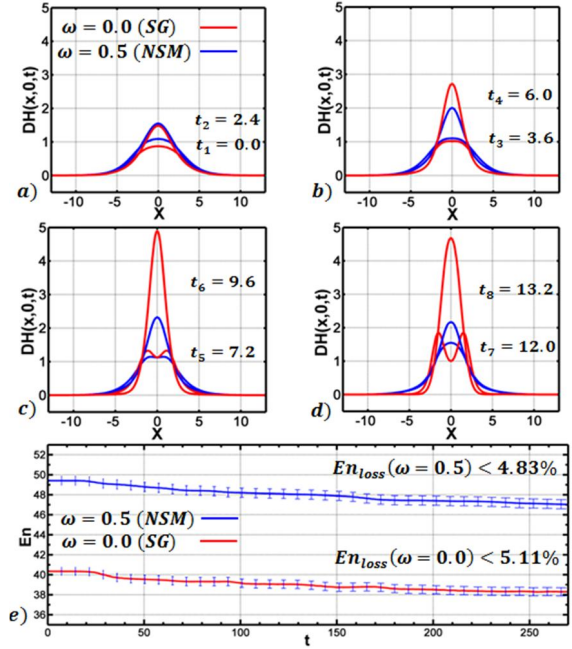

**Fig. 3.** Evolution of the breather solutions (10) in O(3) NSM (8) (at $\omega = 0.5$) and SG (9) (at $\omega = 0.0$). The energy density (DH) in the section $(x, y_0 = 0)$ at $D(v_{t_0}) = \text{sh}(4.38)$: **a)** $t \in [0.0, 2.4]$; **b)** $t \in [3.6, 6.0]$; **c)** $t \in [7.2, 9.6]$; **d)** $t \in [12.0, 13.2]$; **e)** the energy integral of breathers (10) of model (8) at $\omega = 0.5$ and $\omega = 0.0$ (SG). Simulation time: $t \in [0.0, 270.0]$.

Breather field (10) of SG equation (9) with respect to the case of O(3) NSM (8) for $t > 0.0$ has a strong gradient in the central part, as well as large values of amplitude oscillatory dynamics of central points (Fig.3b-d). However, numerical and analytical calculations show that the breather field (10) in the case $\omega = 0.5$ (breathers O(3) NSM) has non-zero values in the wider gradient field and has a relatively large value of the energy integral (En) (Fig.3e) and the bond energy (Fig.3a-d). Moreover, the presence of the rotation of isotopic spin $S(s_1, s_2, s_3)$ in space of sphere $S^2$ (in this case, $\omega = 0.5$) led to some increase in stability of breathers (10) of O(3) NSM.

In the numerical simulation during 45000 iteration cycles ($t \in [0.0, 270.0]$) total energy loss of breathers (10) of the O(3) NSM is:

$$En_{loss}(\omega = 0.5) \approx 4.823\%,$$

which is about 6% less than similar losses breathers of SG equation (Fig.3e):

$$En_{loss}(\omega = 0.0) \approx 5.109\%.$$

### IV. Moving breathers

Properties of Lorentz invariance of O(3) NSM also provides a model of moving solutions. Fig.4 shows the energy density (DH) of the breather (10) of (2+1)-dimensional O(3) NSM (8) for the cases $\omega = 0.0$ (abc) and $\omega = 0.5$ (def), for a given by Lorentz transformation the initial speed $v_{t_0} \approx 0.7071$ ($t_0 = 0.0$). As seen in Fig.4g the moving breather remains stable, and the energy loss by radiation at $t = 45.0$ is $En_{loss} \approx 3.3731\%$ and $En_{loss} \approx 3.5401\%$ for cases $\omega = 0.0$ and $\omega = 0.5$ respectively.

Recall that in our researches on the boundaries of simulation area (in this case: $L_m[3001 \times 3001]$) are inserted special conditions, which absorb excess energy radiated by oscillatory soliton in the form of linear perturbation waves. As in the stationary case, contour projection of energy density (DH) of moving breathers show that in case of $\omega = 0.5$ breathers have a wider non-zero gradient field (Fig.4). This explains the relatively large value of (technical) energy loss of numerical moving breathers when $\omega \neq 0.0$ at the simulation on the small area.

Now let's discuss some of the features of changes of moving breather speed.

In [12] in the study of interaction of one-dimensional breathers of O(3) NSM, was found that speed $v_t$ (at $t > 0.0$) of moving breathers always less than their initial given speed: $v_{t_0} > v_t$ ($t_0 = 0.0$). Whereas, for example, in the study of motion of one-dimensional topological solitons of kink/antikink type always preserved equality: $v_{t_0} \equiv v_t$. Obviously, in the case of breathers, some part of energy absorb by their typical oscillating dynamics.

In Fig.4b (in case $\omega = 0.0$) the breather at $t = 30.0$ passes a distance equal to $s \approx 8$ units (average speed: $v_{t=30} \approx 0.2666$) and the total loss of speed equal to $v_{loss} \approx 0.623\%$. At $t = 45.0$ breather passes a distance equal to $s \approx 12$ units (Fig.4c) and thus the average speed of the moving breather (at $v_{t_0} \approx 0.7071$) is preserved: $v_t \approx 0.2666$ (at $t > 0.0$ and $\omega = 0.0$).

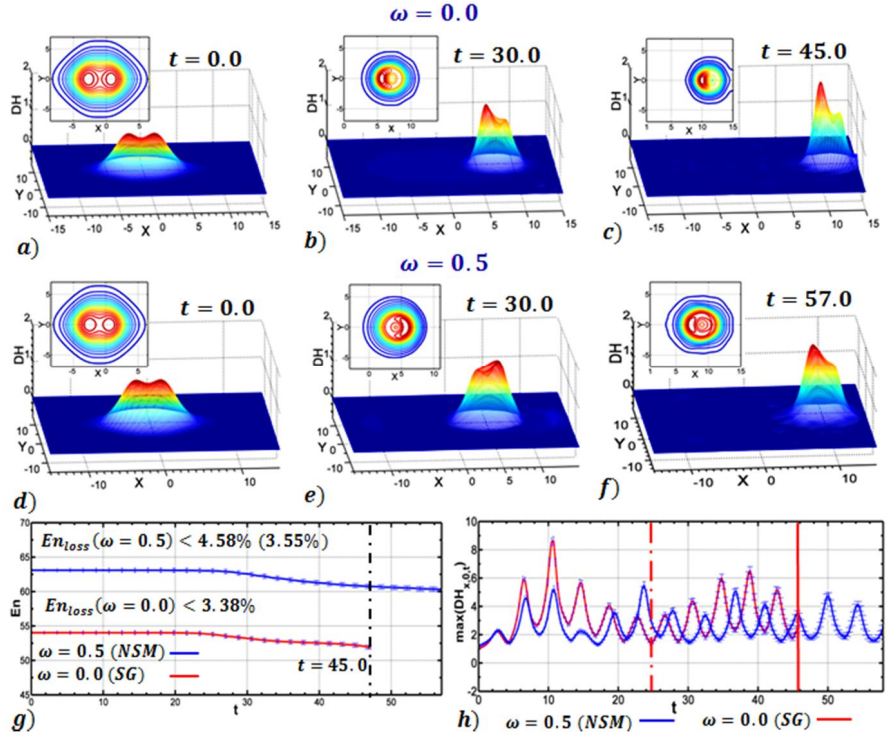

**Fig. 4.** Evolution of the energy density (DH) of moving breather (10) in O(3) NSM (8) with $D(v_{t_0} \approx 0.7) = \text{sh}(4.38)$. For $\omega = 0.0$: **a)** $t = 0.0$; **b)** $t = 30.0$; **c)** $t = 45.0$. For $\omega = 0.5$ **d)** $t = 0.0$; **e)** $t = 30.0$; **f)** $t = 57.0$. For $\omega = 0.0$ and $\omega = 0.5$: **g)** the values of the energy integral (En); **h)** the maximum values of the energy density (DH) in the section $(x, y_0)$. Simulation time: $t \in [0.0, 57.0]$.

On Fig.4e the breather (10) in the case of $\omega = 0.5$ at $t = 30.0$ passes a distance equal to $s \approx 5$ units (average speed: $v_{t=30} \approx 0.1666$) and total loss of speed is equal: $v_{loss} \approx 0.7644\%$. At $t = 57.0$ (for $v_{t_0} \approx 0{,}7071$) breather passes a distance equal to $s \approx 10$ units

(Fig.4f) and thus, at the $t \in [30.0, 57.0]$ we are seeing an increase in the average speed of the moving breather to $v_{t=57} \approx 0.1754$ ($\approx 5.3\%$). Uneven values of average speed of breathers (10) of O(3) NSM (8) in this case (before and after $t = 30.0$) can be explained by the influence of additional rotation dynamics of isotopic spin $S(s_1, s_2, s_3)$ in the space of sphere $S^2$ (in this case, $\omega = 0.5$).

Analysis of the behavior of maximum values of energy density (DH) of moving breathers (10) of the model (8) in the planar section $(x, y_0)$ is shown in Fig.4h: $\max(DH(x, 0, t))$. In the case of $\omega = 0.0$ (breather of SG equation (9)) similar to the stationary case, that shown in Fig.1d can be roughly divided into two wave packets ($t_1 \approx (0.0, 25.0)$, $t_2 \approx (25.0, 50.0)$). In the case of $\omega = 0.5$ (breather of NSM) a point $\max(DH(x, 0, t))$ oscillates with relatively lower average values of frequency and amplitude (Fig.4h). In particular, at the research of speed of the moving breather (10) (at $\omega = 0.5$) a significant reduction ($v_t \to 0.0$) of its values in the interval $t_{v_t} \approx (13.0, 17.0)$ is revealed. These factors explain detected above the speed difference of moving breather (10) before and after the time $t = 30.0$ in the case of $\omega = 0.5$.

## V. Conclusion

Models of stationary and moving breathers of (2+1)-dimensional O(3) NSM obtained in this paper indicate their stability at the different values of the initial speed ($v_{t_0} \leq 0.7071$) of their movement and the rotation frequency ($\omega_{t_0} \leq 0.5$) of vector of isotopic spin $S(s_1, s_2, s_3)$ in $S^2$. Presence of the additional rotation $S$ (in form: $\varphi = \varphi_0 + \omega\tau$, $\omega > 0$) leads to certain dissipation of dynamics of internal degrees of freedom and the energy density (DH) breathers (10) and to increase of their energy integral (En) and bond energy. The results obtained in this paper – stable stationary and moving breather solitons of (2+1)-dimensional O(3) NSM enable the studies of the dynamics of their interaction, which can lead to manifestation of their special particle-like properties [17].

## Acknowledgments

The author is very grateful to Prof. Kh. Kh. Muminov and Prof. B. A. Malomed for useful discussions.

**Appendix**

Evolution of the energy density (DH) of breather solutions (10) of O(3) NSM (8) in meridian section (ω = 0.0) of isotopic space of the sphere $S^2$ (breather of SG equation (9)).

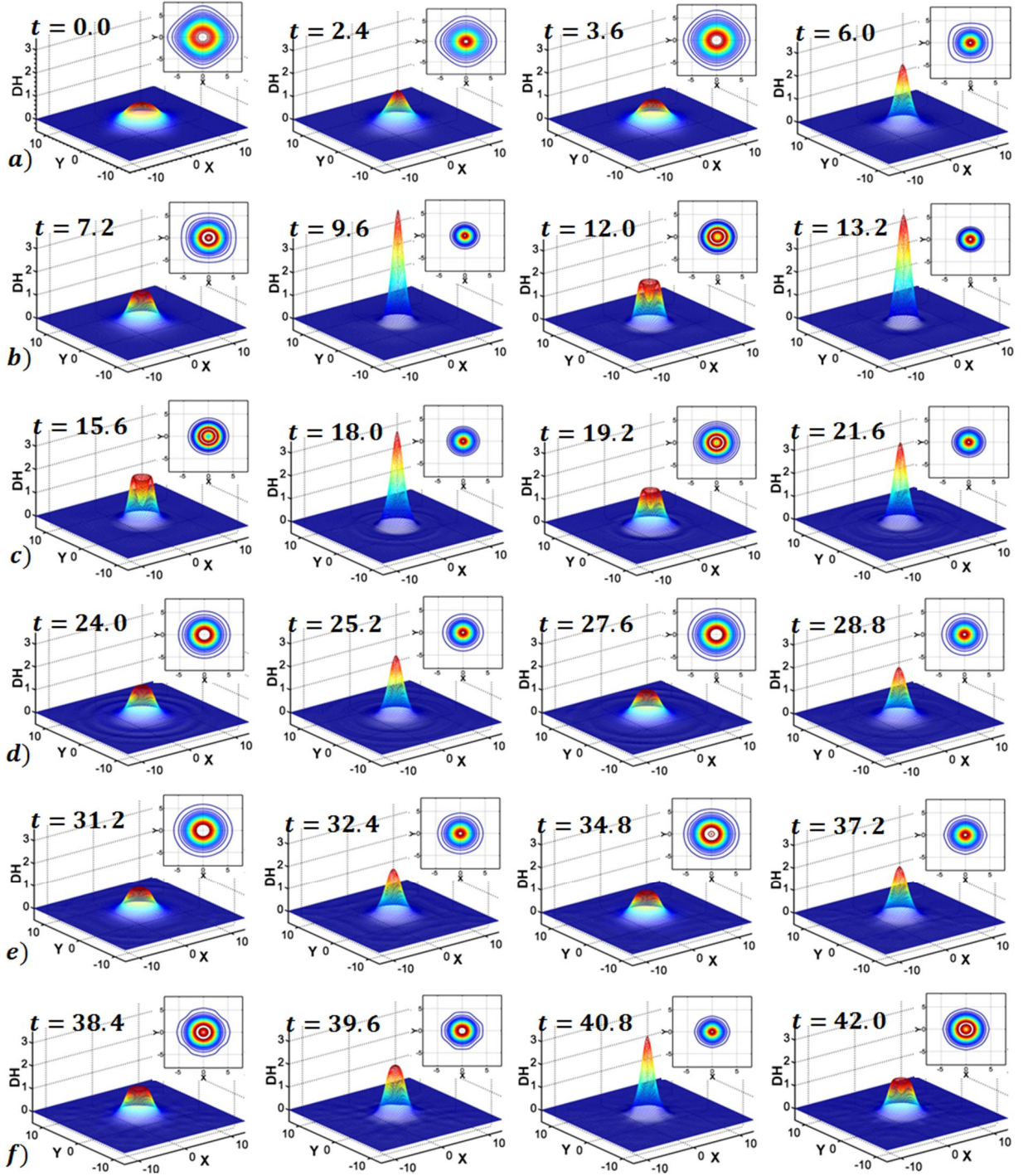

**Fig. A.1.** The energy density (DH) and its contour projection at $D(v_{t_0}) = \text{sh}(4.38)$. The first wave packet: **a)** $t \in [0.0, 6.0]$; **b)** $t \in [7.2, 13.2]$; **c)** $t \in [15.6, 21.6]$; **d)** $t \in [24.0, 28.8]$; **e)** $t \in [31.2, 37.2]$. Second wave packet (part): **f)** $t \in [38.4, 42.0]$. The total simulation time: $t \in [0.0, 270.0]$.

Evolution of the energy density (DH) of breather solutions (10) of O(3) NSM (8) at $\omega = 0.5$.

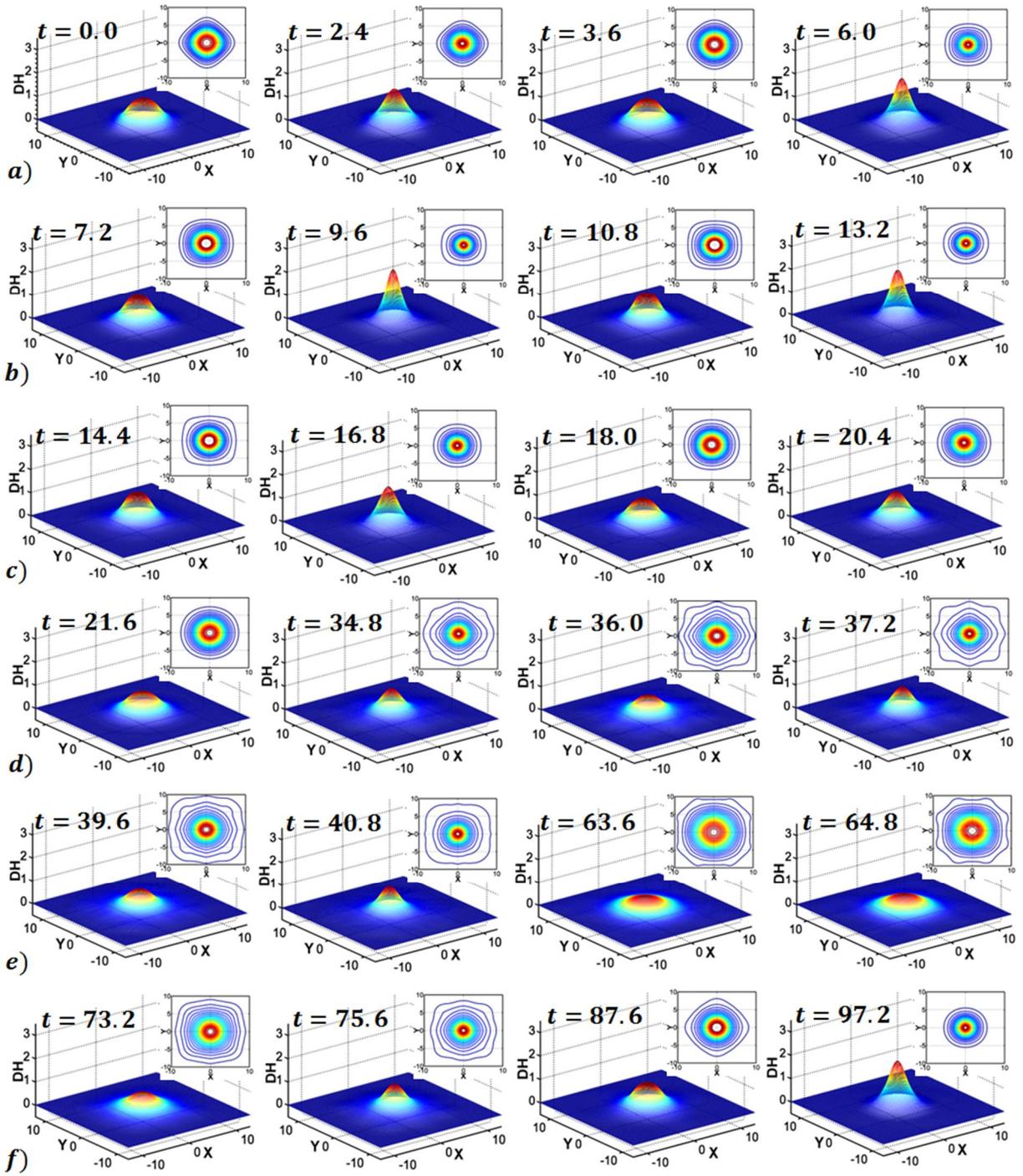

**Fig. A.2.** The energy density (DH) and its contour projection at $D(v_{t_0}) = \text{sh}(4.38)$. The first wave packet: **a)** $t \in [0.0, 6.0]$; **b)** $t \in [7.2, 13.2]$; **c)** $t \in [14.4, 20.4]$; **d)** $t \in [21.6, 37.2]$. Second wave packet (part): **e)** $t \in [39.6, 64.8]$; **f)** $t \in [73.2, 97.2]$. The total simulation time: $t \in [0.0, 270.0]$.